\documentclass[aps,prb,twocolumn,showpacs,amsmath,amssymb,superscriptaddress]{revtex4}

\usepackage{graphicx}
\usepackage{dcolumn}
\usepackage{bm}

\begin{document}

\draft

\title{The effect of screening long-range Coulomb interactions on the metallic behavior in two-dimensional hole systems}

\author{L.H. Ho}
 \email{laphang@phys.unsw.edu.au}
\affiliation{School of Physics, University of New South Wales, Sydney NSW 2052, Australia}
\affiliation{CSIRO Materials Science and Engineering, PO Box 218, Lindfield NSW 2070, Australia}

\author{W.R. Clarke}
\affiliation{School of Physics, University of New South Wales, Sydney NSW 2052, Australia}

\author{A.P. Micolich}
\affiliation{School of Physics, University of New South Wales, Sydney NSW 2052, Australia}

\author{R. Danneau}
	\altaffiliation[Present Address: ]{Low Temperature Laboratory, Helsinki University of Technology, PO Box 3500, 02015 TKK, Finland}
\affiliation{School of Physics, University of New South Wales, Sydney NSW 2052, Australia}

\author{O. Klochan}
\affiliation{School of Physics, University of New South Wales, Sydney NSW 2052, Australia}

\author{M.Y. Simmons}
\affiliation{School of Physics, University of New South Wales, Sydney NSW 2052, Australia}

\author{A.R. Hamilton}
\email{Alex.Hamilton@unsw.edu.au}
\affiliation{School of Physics, University of New South Wales, Sydney NSW 2052, Australia}

\author{M. Pepper}
\affiliation{Cavendish Laboratory, University of Cambridge, Cambridge CB3 0HE, United Kingdom}

\author{D.A. Ritchie}
\affiliation{Cavendish Laboratory, University of Cambridge, Cambridge CB3 0HE, United Kingdom}

\date{\today}
\begin{abstract}
We have developed a technique utilizing a double quantum well
heterostructure that allows us to study the effect of a nearby
ground-plane on the metallic behavior in a GaAs two-dimensional
hole system (2DHS) in a single sample and measurement cool-down,
thereby maintaining a constant disorder potential. In contrast to
recent measurements of the effect of ground-plane screening of the
long-range Coulomb interaction in the insulating regime, we find
surprisingly little effect on the metallic behavior when we change
the distance between the 2DHS and the nearby ground-plane.
\end{abstract}


\maketitle

The ground state of a 2D electron or hole system is determined by
the ratio $r_{s}$ of the inter-particle Coulomb energy to the
kinetic energy, with the regimes $r_{s} \sim 0$, $r_{s} \sim 10$,
and $r_{s} \sim 100$ corresponding to the gas, liquid and solid
phases of the 2D system. Recently, much effort has gone into
studying the phase diagram of 2D systems, focussing on the role of
interactions and disorder \cite{Abrahams01}. An interesting
approach is to use an adjacent metallic ground-plane to screen the
long-range Coulomb interactions, and thereby study how the
length-scale of the Coulomb interactions controls the ground state
of the 2D system. Ground-plane screening was used to probe the
effect of Coulomb interactions in the melting of the Wigner
crystal state of electrons on liquid helium \cite{Jiang88,
Mistura97}. More recently, Huang \emph{et al.} \cite{Huang06} have
used a nearby ground-plane to study the insulating state in an
ultra-low density 2D hole system (2DHS), and find that the
insulating behaviour is strongly affected by screening the Coulomb
interaction. It would be interesting to perform a corresponding
study for the metallic behavior observed in a dilute 2DHS.
However, this presents a significant technical challenge because
on one hand, the higher hole density $p$ in the metallic regime
requires that the distance $d$ between the ground-plane and the
2DHS be quite small ($d \sim 50$ nm) to achieve effective
screening, and on the other, 2DHSs of sufficient quality to
observe the metallic behavior are typically buried deep ($> 100$
nm) in the semiconductor.

In this paper, we report a study of the influence of
ground-plane screening on the metallic behavior in a 2DHS. We find
that screening the long-range Coulomb interactions has a
relatively small effect in the metallic regime, in contrast to the
insulating behavior
\cite{Huang06}. Our findings suggest that hole-hole screening
within the 2DHS already significantly reduces long-range Coulomb
interactions in the metallic regime, consistent with the view that
the metallic behavior arises due to temperature-dependent
screening of the Coulomb interactions between holes and charged
impurities \cite{Gold90,DasSarma00,Zala01}. To overcome the
technical challenge of studying the ground-plane screening in the
metallic regime, we have used a heterostructure featuring two
quantum wells buried $\sim 300$ nm beneath the semiconductor
surface and separated by only $50$ nm, such that the 2DHS in the
upper quantum well acts as the nearby ground-plane for the 2DHS in
the lower quantum well. Using a metal surface-gate, we can deplete
the upper 2DHS, making the surface-gate become the ground-plane,
and thus increasing the distance $d$ to the ground-plane by a
factor of $\sim 7$. This is a major advantage as $d$ can be
changed in a single device and measurement cool-down, thereby
keeping the disorder relatively constant, unlike previous studies.

\begin{figure}[tbph]
\includegraphics[width=8cm]{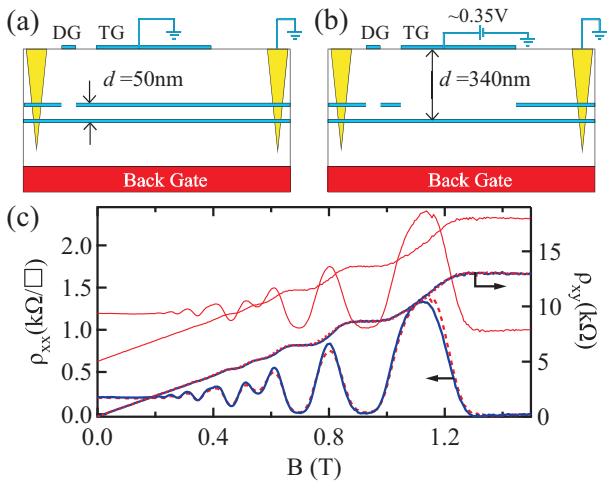}
\caption{\label{fig:SdeH-comparison} (Color online) Schematic of the technique
used to vary the distance $d$ to the ground plane, where the upper
2DHS is (a) occupied ($d = 50$ nm) or (b) depleted ($d = 340$ nm).
(c) A comparison of the longitudinal ($\rho_{xx}$ - left axis) and
Hall ($\rho_{xy}$ - right axis) resistance of the lower 2DHS at $p
= 6.6 \times 10^{10}$ cm$^{-2}$ with $V_{BG}=0$ V for $d = 340$
nm (solid blue lines) and $d = 50$ nm (dashed red lines). The
arrows indicate the relevant axes. The data for $d = 50$ nm has
been duplicated and offset upwards by 1 k$\Omega/\square$ in
$\rho_{xx}$ and 5 k$\Omega$ in $\rho_{xy}$ (solid red lines) for
clarity, since the data for the two $d$ values overlap.}
\end{figure}

Our approach is illustrated in Fig. 1(a/b), and utilizes a double
quantum well (DQW) structure featuring a doped semiconductor
back-gate, an overall top-gate (TG) and a set of depletion gates
(DG) adjacent to each AuBe ohmic contact. The top and back gates
allow the hole densities $p$ in the two 2DHSs to be controlled.
The ohmic contacts connect to both QWs and the depletion gate is
used to sever all connections to the upper 2DHS with the exception
of the drain contact, via which the upper 2DHS is grounded (see
Fig. 1(a)). Hence current passes only via the lower 2DHS. In this
experiment, we measure the temperature dependent resistivity
$\rho(T)$ of the lower 2DHS when it is separated from a nearby
ground plane by $d = 50$ nm (Fig. 1(a)) and $d = 340$ nm (Fig.
1(b)). For the $d = 50$ nm case, the top-gate is set to 0 V,
resulting in a 2DHS with a density of approximately
$1\times10^{11}$ cm$^{-2}$ in the upper QW, which acts as a ground
plane. For the $d = 340$ nm case, the top gate is set to $\sim
+0.35$ V, depleting the upper 2DHS in regions directly beneath the
top gate. Despite being positively biased, the top-gate is still
an equipotential, and hence acts as a ground-plane when the upper
2DHS is depleted. Combining this operating principle with
appropriate adjustments to the various gate biases, we can switch
between the two $d$ values while keeping the density in the lower
2DHS constant, and repeat the experiment at different densities.

The device is fabricated on a (311)A AlGaAs/GaAs DQW
heterostructure, featuring two 20nm GaAs QWs separated by a 30nm
thick Al$_{0.3}$Ga$_{0.7}$As barrier that ensures there is no
tunneling between the two 2DHSs ($\Delta _{SAS}\ll 0.1$ K). The
heavily-doped (311)A substrate acts as the back-gate and is
separated by 3 $\mu$m from the lower QW. The device is fabricated
into a Hall bar structure oriented along the high-mobility
$[\overline{2}33]$ direction. Transport measurements were performed
in a Kelvinox 100 dilution refrigerator, with a 20 mK base
temperature. Standard four-terminal low-frequency a.c lock-in
techniques were used with a constant excitation voltage of 20 $\mu$V
applied at $f = 4$ Hz.

In Fig. 1(c) we present magnetotransport measurements in the two
device configurations presented in Fig. 1 (a/b) to demonstrate
that we can maintain a constant $p$ in the lower 2DHS while
changing the ground-plane distance from $d = $ 50 nm (dashed red
trace) to $d = $ 340 nm (solid blue trace). Due to the close
overlap of these two traces, we have reproduced the $d = $ 50 nm
traces and offset them vertically for clarity (solid red lines).
The Hall slope ($\rho_{xy}$ - right axis) and the period of the
Shubnikov-de Haas (SdH) oscillations ($\rho_{xx}$ - left axis)
show that the density remains constant to within $1\%$ despite
changing $d$ by a factor of $\sim$7.

\begin{figure}
\includegraphics[width=8cm]{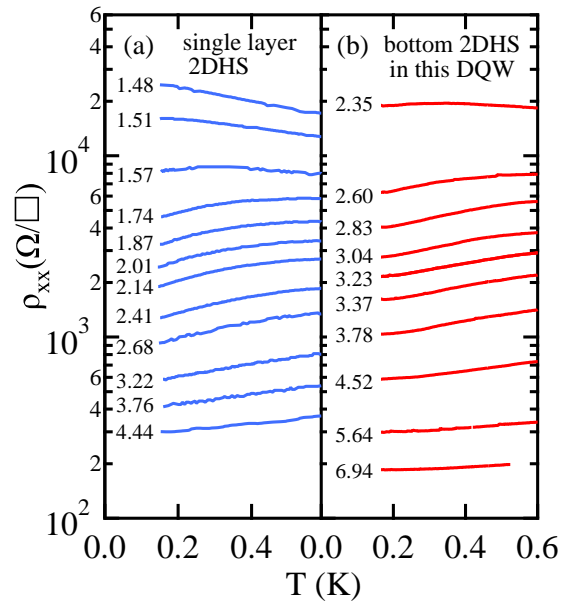}
\caption{\label{fig:T402vsT483} (Color online) A comparison of the temperature
dependent longitudinal resistivity $\rho_{xx}(T)$ vs temperature $T$
for: (a) a previously measured 2DHS in a similar single QW
heterostructure \cite{Hamilton01}, and (b) the lower 2DHS of our
device when the upper 2DHS is depleted ($d = 340$ nm). The numbers
beside each trace indicate the hole density in units of $10^{10}$
cm$^{-2}$.}
\end{figure}

We commence our study by comparing the temperature dependent
resistivity $\rho_{xx}(T)$ measured for $d = 340$ nm with that
obtained in a prior study using a similar p-GaAs heterostructure
containing a single QW \cite{Hamilton01}. This allows us to confirm
that the depleted upper 2DHS in our device does not significantly
alter the results obtained for large $d$. In Fig. 2(a) we show
$\rho_{xx}(T)$ data obtained in \cite{Hamilton01}, where a clear
metal-insulator transition is observed. Insulating behavior
($\frac{\partial\rho}{\partial T} < 0$) is observed at low
densities, with a transition to metallic behavior
($\frac{\partial\rho}{\partial T} > 0$) occurring at $p \approx 1.57
\times 10^{10}$ cm$^{-2}$. With further increases in $p$, the
metallic behavior weakens again, as expected because the hole-hole
interactions weaken with increasing $p$. Very similar behavior is
obtained in our device for $d = 340$ nm, as shown in Fig. 2(b). At
high densities, weak metallic behavior is observed, which becomes
stronger as $p$ is decreased. We are unable to measure in the
insulating regime because our device becomes unstable when the
top-gate bias is too high, however this is not an impediment to the
experiments presented here.

\begin{figure}
\includegraphics[width=8cm]{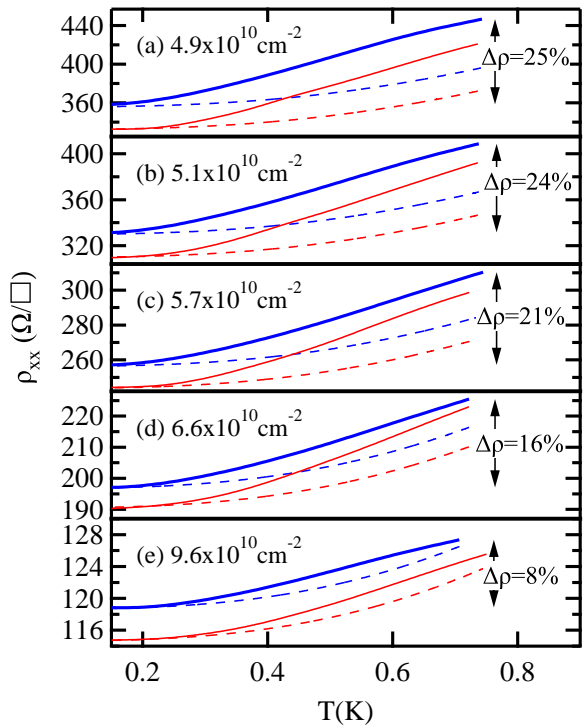}
\caption{\label{fig:tdeps} (Color online) (a)-(e) The measured resistivity
$\rho_{xx}(T)$ vs temperature $T$ for $d = 50$ nm (thin red lines)
and $d = 340$ nm (thick blue lines) for five different densities
$p$. The back-gate bias was the same for the two $d$ values at
each density. The dashed lines show the calculated phonon
contributions (parameter free fits of Karpus' theory) in each
case. The percentage change in resistivity from $150$ mK to $0.7$
K for $d = 340$ nm is shown for each panel.}
\end{figure}

In Fig. 3 we present $\rho_{xx}(T)$ measured with $d =  340$ nm
(thick lines) and $d = 50$ nm (thin lines) at five different
densities deep in the metallic regime. Moving down through the
panels in Fig. 3, the metallic behavior weakens with increasing
$p$, as indicated by the percentage change in resistivity $\Delta
\rho$ over the measured temperature range.  The most obvious
effect of reducing $d$ to 50 nm is to offset $\rho_{xx}(T)$ to a
lower resistivity, likely caused by the upper 2DHS screening the
lower 2DHS from the long-range impurity scattering from the upper
modulation doping layer.

The resistivity $\rho_{xx}(T)$ is predominantly composed of three
parts: a temperature-independent Drude contribution $\rho_{0}$ due
to impurity scattering, and two temperature-dependent
contributions due to phonon scattering $\rho_{ph}(T)$ and
hole-hole interactions. We calculate $\rho_{ph}(T)$ using Karpus'
theory \cite{Karpus90}, as done previously for 2D hole systems
\cite{Prosku02,Yasin05}. The Drude contribution $\rho_{0}$ is then
obtained by extrapolating $\rho(T) - \rho_{ph}(T)$ to $T = 0$. The
calculated phonon contributions for each $d$ and $p$ are presented
as dashed lines in Fig. 3. At the highest density, where
interactions are weakest, $\rho_{ph}(T)$ is a remarkably good fit
to the measured $\rho(T)$ considering there are no fitting
parameters, giving confidence in our application of
Ref.~\onlinecite{Karpus90} to calculating the phonon contribution.
In contrast, at lower densities, there is a significant
temperature dependence that cannot be described by phonon
scattering alone, which we attribute to hole-hole interactions
(screening).

Following Proskuryakov \emph{et al.}, we isolate the contribution
of the Coulomb interactions to the conductivity $\sigma_{int}(T)$
by subtracting the phonon and Drude contributions using
$\sigma_{int}(T) = [\rho(T) - \rho_{ph}(T)]^{-1} - \rho_{0}^{-1}$
\cite{Prosku02}. In Fig. 4 (a), we plot $\sigma_{int}(T)$ vs $T$
corresponding to the $d = 50$ nm trace in Fig. 3(a), and find an
approximately linear dependence of $\sigma_{int}$ on $T$, with a
negative slope corresponding to metallic behavior. It is common to
interpret such data using the theory of Zala \emph{et al.}
\cite{Zala01} to obtain the Fermi liquid interaction parameter
$F_{0}^{\sigma}$. This is normally achieved by fitting
$\sigma_{int}$ vs $T$ in the linear regime (i.e., $T << T_{F}$,
where $T_{F}$ is the Fermi temperature) to obtain the normalized
slope $d(\sigma_{int}/\sigma_{0})/d(T/T_{F})$, which is related to
$F_{0}^{\sigma}$ by $d(\sigma_{int}/\sigma_{0})/d(T/T_{F}) = 1 +
[3F_{0}^{\sigma}/(1 + F_{0}^{\sigma})]$ \cite{Noh03}. However,
recent studies have shown that the theory of Zala \emph{et al.}
\cite{Zala01}, which assumes a short-range scattering potential,
does not extract the correct values of $F_{0}^{\sigma}$ for
modulation-doped samples where long-range scattering from remote
ionized impurities is significant \cite{Clarke07}. Hence in this
paper, we focus on the slope
$d(\sigma_{int}/\sigma_{0})/d(T/T_{F})$ as a measure of the
strength of the metallic behavior, henceforth called the
`metallicity', and simply note that if we fit Zala's theory to our
data, we obtain $F_{0}^{\sigma}$ values similar to those
previously obtained in modulation-doped GaAs 2D hole systems by
Proskuryakov \emph{et al.} \cite{Prosku02}.

\begin{figure}
\includegraphics[width=8cm]{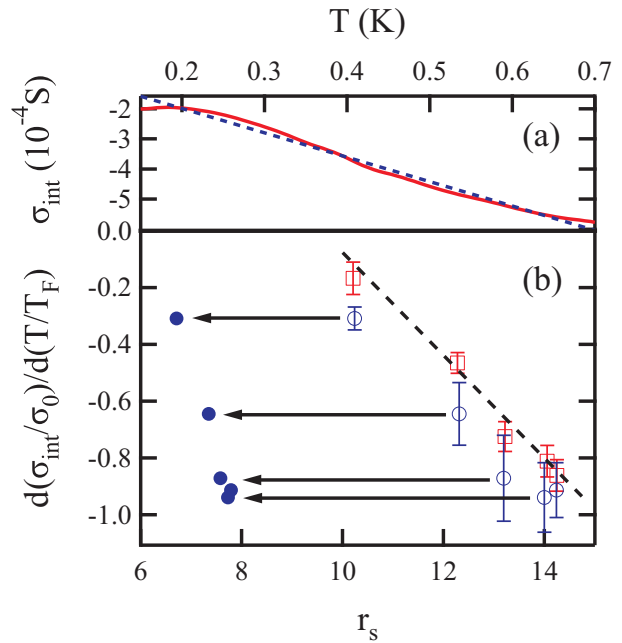}
\caption{\label{fig:F0vsRs} (Color online) (a) $\Delta \sigma$ vs $T$ for $p=4.9
\times 10^{10} cm^{-2}$ and $d=50$nm. The data is the thin solid
line and the linear fit is the thick dashed line. The error bars
indicate the effect of varying the temperature range of the linear
fits used to obtain $d(\sigma_{int}/\sigma_{0})/d(T/T_{F})$.
Although the values change slightly, the observed trends are
unaffected. (b) A plot of the metallicity
$d(\sigma_{int}/\sigma_{0})/d(T/T_{F})$ vs $r_{s}$ for $d = 340$
nm (open squares) and $d = 50$ nm (open circles), and
$r_{s}^{dipole}$ for $d = 50$ nm (closed circles). The dashed line
is a guide to the eye for the trend exhibited by the $d = 340$ nm
data vs $r_{s}^{bare}$ (open squares).}
\end{figure}

We now directly analyze the effect of changing the ground-plane
separation $d$, and thus the length-scale of the Coulomb
interactions, on the metallic behavior. To do this, we obtain the
metallicity $d(\sigma_{int}/\sigma_{0})/d(T/T_{F})$ for each trace
in Fig. 3 by taking a linear fit to the corresponding
$\sigma_{int}$ vs $T$ data (Fig. 4(a)). We limit our fitting to
the range $150$ mK $< T < 0.7$ K. The lower limit is the minimum
hole temperature in our experiment (obtained following
\cite{McPhail04}) whilst the upper limit ensures that $T <<
T_{F}$. The relative strength of the hole-hole Coulomb
interactions is characterized by $r_{s}$, which in the absence of
a nearby ground-plane, takes the form $r_{s}^{bare} =
(m^{*}e^{2})/(4\pi^{3/2}\hbar^{2}\epsilon\sqrt{p})$, where
$\epsilon$ is the dielectric constant. In Fig. 4(b), we plot the
metallicity $d(\sigma_{int}/\sigma_{0})/d(T/T_{F})$ as a function
of $r_{s}^{bare}$ for $d = 340$ nm (open squares) and $d = 50$ nm
(open circles). In each case
$d(\sigma_{int}/\sigma_{0})/d(T/T_{F})$ is negative, indicating
metallic behavior. As Fig. 4(b) shows, there is little change in
the metallicity as $d$ is reduced from 340 nm to 50 nm at constant
density, but the metallic behavior weakens with increasing density
(i.e., decreasing $r_{s}$), as expected. The presence of the
ground-plane introduces negative image-charges that cause the
holes to interact as though they were dipoles rather than single
positive charges, resulting in a Coulomb potential that falls off
as $\sim 1/r^{3}$ instead of $1/r$ for $r \geqslant 2d$
\cite{Huang06}. This leads to an additional term in $r_{s}$ that
accounts for the Coulomb interaction between the hole's
image-charge and another hole in the 2DHS, giving
\cite{Widom88,Ho08}:
$$r_{s}^{dipole} = \frac{m^{*}e^{2}}{4\pi^{2}\hbar^{2}\epsilon
p}[(\pi p)^{1/2}-(\frac{1}{\pi p}+d^{2})^{-1/2}]$$

The additional term in Eqn (1) reduces the effective $r_{s}$ as
$d$ becomes comparable to the hole-hole separation \cite{Huang06} $a = 2(\pi
p)^{1/2}$, with $a$ ranging from $36 - 51$ nm for
the data in Fig. 3. Thus, whilst for $d = 340$ nm the
effective reduction in $r_{s}$ is quite small ($< 8\%$), it is
quite substantial for $d = 50$ nm, as demonstrated in Fig. 4(b),
where we have re-plotted the  metallicity for $d = 50$ nm vs.
$r_{s}^{dipole}$ (solid circles).

We now consider the effect of the ground plane on the metallicity.
The dashed line in Fig. 4(b) shows the metallicity of the
unscreened $d = 340$ nm data tends to decrease as $r_s$ is
reduced. If the metallicity depends only on $r_{s}$, for example,
if it is due to long-range Coulomb interactions \cite{Zala01},
then when the effective $r_{s}$ is decreased by reducing $d$ from
340 nm to 50 nm, the metallicity should also be reduced. When
plotted against $r_{s}^{dipole}$, the metallicity for the $d = 50$
nm data (solid circles) should lie in the upper left corner of
Fig. 4(b), not down in the lower left of the figure. In fact
rather than being reduced, the $d=50$ nm metallicity is actually
slightly greater than for $d = 340$ nm, suggesting that the
ground-plane has almost no effect on the metallic behavior. This
is in contrast to recent experiments on the screening of
long-range Coulomb interactions in the insulating regime
\cite{Huang06}, where single layer 2DHS samples with $d=$ 250, 500
and 600 nm were examined. The hopping conductivity was fitted to
the form $\sigma/(e^2/h)=G_{0} + (T/T_{0})^\alpha$, and the
exponent $\alpha$, which is related to the strength of the
insulating behavior,  was found to be very sensitive to \emph{d}, and  uniquely determined by the
ratio $d/a$, which was changed from 1.1 to 5 [4]. In our experiment, $d/a$ 
changes by almost an order of magnitude, from 2 to 19, 
yet we find that the metallicity is primarily determined by $a$ alone.


One possible explanation for the contrasting behaviour is that in
the metallic regime the long-range Coulomb interactions are
already screened by the other holes in the 2DHS, significantly
diminishing the effect of the ground-plane.
In the insulating regime, this screening by other holes is much
weaker, giving the ground-plane significant effect. This is
consistent with recent studies of the compressibility of 2D
systems \cite{Allison06}, which found that the Thomas-Fermi
screening radius increases by over an order of magnitude when
going from the metallic to insulating regime.

We finish with some final comments on the experiment. First, it is
particularly interesting to note that our findings are consistent
with the metallic behavior being due to the temperature-dependent
screening of the hole-impurity Coulomb interactions
\cite{Gold90,DasSarma00,Zala01}.
Second, because the screening layer for $d = 50$ nm is a nearby
2DHS, we have estimated the possible contribution of Coulomb drag
to $\Delta \rho(T)$, since this can be a significant effect in DQW
structures \cite{Pillar05}. Due to the relatively large QW spacing
and high densities, we calculate the drag contribution to $\Delta
\rho(T)$ to be $<1\%$ in our device. Finally, in addition to
theoretical calculations~\cite{Ho08}, we have performed
experiments to check that the upper 2DHS density is sufficient to
act as a proper screening layer. We find no change in the metallic
behavior when the upper 2DHS density is varied~\cite{Ho07},
confirming that the upper 2DHS acts as an effective ground-plane.


This work was funded by Australian Research Council (ARC) and the
EPSRC. L.H.H. acknowledges support from the UNSW and the
CSIRO. O.K. acknowledges support from the UNSW. M.Y.S,
A.R.H, A.P.M. and R.D. acknowledge support from the ARC. We thank O.P. Sushkov for
helpful discussions and J. Cochrane for technical support.


\begin{thebibliography}:

\bibitem{Abrahams01} B.L. Altshuler, D.L. Maslov and V.M. Pudalov, Physica E {\bf9}, 2, (2001); S. V. Kravchenko and M. P. Sarachik, Rep. Prog. Phys., {\bf67}, 1, (2004)


\bibitem{Jiang88} H.-W. Jiang and A. J. Dahm, Surf. Sci. {\bf196}, 1 (1988).

\bibitem{Mistura97} G. Mistura, T. G\"{u}nzler, S. Neser, and P. Leiderer, Phys. Rev. B {\bf56}, 8360 (1997).

\bibitem{Huang06} J. Huang, D. S. Novikov, D. C. Tsui, L. N. Pfeiffer and K. W. West, Cond-mat/0610320.

\bibitem{Gold90} A.Gold, Phys. Rev. B {\bf41}, 8537 (1990).

\bibitem{DasSarma00} S. Das Sarma and E.H. Hwang, Phys. Rev. B {\bf61}, R7838 (2000).

\bibitem{Zala01} G. Zala, B. N. Narozhny, and I.L. Aleiner, Phys. Rev. B {\bf64}, 214204 (2001).

\bibitem{Hamilton01} A. R. Hamilton, M. Y. Simmons, M. Pepper, E. H. Linfield, and D. A. Ritchie, Phys. Rev. Lett. {\bf 87}, 126802 (2001).

\bibitem{Prosku02} Y. Y. Proskuryakov, A. K. Savchenko, S. S. Safonov, M. Pepper, M. Y. Simmons, and D. A. Ritchie, Phys. Rev. Lett. {\bf89}, 076406 (2002).

\bibitem{Karpus90} V. Karpus, Semicond. Sci. Tech. {\bf5}, 691 (1990).

\bibitem{Yasin05} C. E. Yasin, T. L. Sobey, A. P. Micolich, A. R. Hamilton, M. Y. Simmons, W. R. Clarke, L. N. Pfeiffer, K. W. West, E. H. Linfield, M. Pepper, and D. A. Ritchie, Phys. Rev. B {\bf72}, 241310(R) (2005).

\bibitem{Noh03} H. Noh, M. P. Lilly, D. C. Tsui, J. A. Simmons, E. H. Hwang, S. Das Sarma, L. N. Pfeiffer, and K. W. West, Phys. Rev. B {\bf68}, 165308 (2003).

\bibitem{Clarke07} W. R. Clarke, C. E. Yasin, A. R. Hamilton, A. P. Micolich, M. Y. Simmons, K. Muraki, Y, Hirayama, M. Pepper and D. A. Ritchie, Nature Physics {\bf4}, 55 (2008).

\bibitem{McPhail04} S. McPhail, C.E. Yasin, A.R. Hamilton, M.Y. Simmons, E.H. Linfield, M. Pepper and D.A. Ritchie, Phys. Rev. B {\bf70}, 245311 (2004).

\bibitem{Widom88} A. Widom and R. Tao, Phys. Rev. B {\bf38}, 10787 (1988).

\bibitem{Ho08} O.P. Sushkov (Private Communication); calculations showing that the 2DHS produces almost exactly the same effect as a metal gate will be published elsewhere.

\bibitem{Pillar05} R. Pillarisetty, H. Noh, E. Tutuc, E. P. De Poortere, K. Lai, D. C. Tsui and M. Shayegan, Phys. Rev. B {\bf71}, 115307 (2005).



\bibitem{Allison06}  G. Allison, E. A. Galaktionov, A. K. Savchenko, S. S. Safonov, M. M. Fogler, M. Y. Simmons, and D. A. Ritchie, Phys. Rev. Lett. {\bf96}, 216407 (2006).

\bibitem{Ho07} L.H. Ho, W.R. Clarke, R. Danneau, O. Klochan, A.P. Micolich, M.Y. Simmons, A.R. Hamilton, M. Pepper and D.A. Ritchie, Physica E {\bf40}, 1700 (2008)

\end{thebibliography}
\end{document}